\input epsfig.sty

\documentstyle{EuroPre}

\def\etal{{\hbox{{\tenit\ et al.\/}\tenrm :\ }}}

\def\And{{\rm and\ }}

\newif\ifboo \boofalse

\begin{document}


\euro{}{}{}{}
\Date{April 1997}
\shorttitle{A. D. RUTENBERG \etal DYNAMICAL MULTISCALING IN QUENCHED SKYRME 
SYSTEMS}

\title{Dynamical Multiscaling in Quenched Skyrme Systems}
\author{A. D. Rutenberg\inst{1}, 
        W. J. Zakrzewski\inst{2} \And
        M. Zapotocky\inst{3} }
\institute{
     \inst{1} Theoretical Physics, University of Oxford, Oxford OX1 3NP, United
Kingdom \\
     \inst{2} Department of Mathematical Sciences, University of Durham, Durham
DH1 3LE, United Kingdom \\
     \inst{3} Department of Physics and Astronomy, University of Pennsylvania, 
Philadelphia, PA 19104, USA}

\rec{}{}

\pacs{
\Pacs{64}{60.Cn}{Order-disorder transitions and statistical mechanics
 of model systems}
\Pacs{11}{27$+$d}{Extended classical solutions; cosmic strings, domain
 walls, texture}
\Pacs{12}{39.Dc}{Skyrmions}
}

\maketitle

\begin{abstract}
Strong dynamical 
scaling violations exist in quenched two-dimensional systems with
vector $O(3)$ order parameters. These systems 
support non-singular topologically stable configurations 
(skyrmions). By tuning the stability of isolated skyrmions to 
expand or shrink, we find dramatic differences in the dynamical multiscaling
spectrum of decaying moments $\langle |\rho|^n \rangle$
of the topological charge density distribution
and in particular in the decay of the energy-density
$\epsilon \sim \langle |\rho| \rangle$. We present a simple two-length-scale
model for the observed exponents in the case when isolated skyrmions expand. 
No such simple model is found when isolated skyrmions shrink.
\end{abstract}


Phase-ordering systems quenched from disordered initial conditions into
an ordered phase typically evolve correlations that dynamically scale
--- they are self-similar in time \cite{Bray94}. Dynamical scaling 
provides a powerful framework for the analysis of phase-ordering 
exponents and correlations. To understand scaling,  
systems that {\em violate} scaling are important to study. 
Logarithmic scaling violations exist in the 
spherical limit of conserved vector systems \cite{Coniglio89}. 
Recently, strong scaling violations
have been found in one-dimensional \cite{Rutenberg95c}
and two-dimensional (2D) \cite{Rutenberg95a,Zapotocky95} 
systems supporting non-singular topological textures (``skyrmions'').

In this paper we investigate the phase-ordering process of the
2D $O(3)$ vector model in more detail, paying particular attention 
to sub-dominant terms in the free energy. These terms control the stability of
individual skyrmions towards shrinking or growing in size. 
Surprisingly, these sub-dominant terms
dramatically change the exponent $\phi_N$ characterizing the decay of
energy in the system.

We restrict ourselves to 
dissipative noise-free ``model-A'' dynamics, $\partial_t \vec{m} = 
-  \delta F/ \delta \vec{m}$, where $|\vec{m}|=1$ is
maintained as a constraint.  Our effective coarse-grained free energy is 
\begin{equation}
\label{EQ:en}
F[\{\vec{m}\}] ={1 \over 8 \pi} \int d^2 {\bf r}\,({\bf\nabla} \vec{m})^2 
+ \pi \theta  \int d^2 {\bf r}\, \rho^2,
\end{equation}
where the topological charge density $\rho$ is
\begin{equation}
\label{EQ:rho}
\rho({\bf r}) 
= \vec{m} \cdot (\partial_x \vec{m} \times \partial_y \vec{m}) / 4 \pi,
\end{equation}
and is locally conserved in continuum systems \cite{Rajaraman82}. 
The term $\pi \theta  \int d^2 {\bf r}\, \rho^2$ in Eqn.~(\ref{EQ:en})
is analogous, when $\theta$ is positive, to the Skyrme term used to stabilize 
skyrmion configurations in high-energy models \cite{Skyrme61}.
At time $t=0$, we randomly orient the unit vector $\vec{m}$ at each site. 
This initial condition corresponds to a quench from $T=\infty$ to $T=0$.

A 2D skyrmion (see, e.g., \cite{Rajaraman82}) 
is a field configuration $\vec{m}({\bf r})$ 
that wraps around the O($3$) order-parameter sphere exactly once.  
It is described by a
spatially extended topological charge $\rho({\bf r})$ that integrates
to $\pm 1$.  Each skyrmion can be characterized by a size, defined, e.g., 
as the half-width of the typically 
bell-shaped spatial distribution of topological charge $\rho({\bf r})$.
In a 2D O($3$) phase-ordering system, which has a mixture of skyrmions
and anti-skyrmions, we can define 
at least three natural length-scales: 
the average skyrmion size $L_T \sim t^{\phi_T}$, 
the average separation of skyrmions
$L_N \sim t^{\phi_N}$, and a length characterizing positive and negative
charge separation --- for example 
the inverse interface density, $L_C \sim t^{\phi_C}$, of $\rho =0$ contours. 
All three length-scales are found to have distinct growth exponents.

The static (minimal energy) configurations of an infinite {\em continuum}
system with a pure quadratic gradient energy ($\theta=0$)
were studied by Belavin and Polyakov (BP) \cite{Belavin75}. 
For any net number of skyrmions in the system, they found a class of 
minimal energy states that are 
degenerate with respect to both the scale and position of the 
individual skyrmions. Essentially, the skyrmions 
in the pure skyrmion (or pure anti-skyrmion) BP configurations do not interact,
and they neither expand nor shrink.

In generic systems, the energy density will also contain 
higher-order gradient terms, arising e.g. from lattice interactions.  
The dominant corrections at late times (and hence large length-scales) 
in quenched systems will come from fourth-order
gradient terms, such as the Skyrme term in Eqn.~(\ref{EQ:en}).
Such terms will destabilize the individual
skyrmions towards expanding or shrinking. 
By adding the Skyrme term in Eqn.~(\ref{EQ:en})
with a sufficiently large amplitude $\theta > \theta_c$, 
it is possible to ensure that all skyrmions expand 
\footnote{Our Skyrme term in Eqn.~(\protect\ref{EQ:en})
does not describe all possible fourth-order gradient terms.
For general fourth-order terms,
the energy of a spherically symmetrical skyrmion of size $R$ 
is $E(t) \sim  1 + A (\theta-\theta_c) R^{-2}$, 
where $A$ is a positive constant.
The rate of change of the energy is $\partial_t E \sim \dot{R}^2$
\cite{Rutenberg95b}. Self-consistency gives $\partial_t {R} \sim - R^{-3}$,
which leads (in our case) to $R(t) \propto [(\theta-\theta_c)(t-t_0)]^{1/4}$.
Skyrmions are also destabilized towards shrinking, with a similar rate,
in the {\it continuum\/} model when a soft-spin potential that allows 
variations in the spin magnitude is used \protect\cite{Rutenberg95a}.}.
In our simulations, we use a $9$-point Laplacian with nearest and
next-nearest neighbor interactions and an 
isotropic fourth-order term. Our leading interactions in Fourier
space are proportional to $(k^2-k^4/24) |\vec{m}_{\bf k}|^2$, 
resulting in a positive $\theta_c$.

Previous work \cite{Rutenberg95a,Zapotocky95,Toyoki93,Bray90} 
has considered systems where isolated skyrmions shrink and 
ultimately unwind through the lattice ($\theta<\theta_c$), thus violating 
charge conservation. For $\theta>\theta_c$, topological charge density
is eliminated solely through the mutual annihilation of regions of positive
and negative $\rho$ --- a process conserving the topological charge.
For $\theta < \theta_c$, skyrmion unwinding and
skyrmion/anti-skyrmion annihilation are competing processes ---
a partially annihilated skyrmion cannot unwind by itself \cite{Zapotocky95}.
We re-analyze our earlier results \cite{Zapotocky95},
and also study systems with $\theta>\theta_c$ using the same 
numerical techniques. We use $\theta=0.012$ in
dimensionless units with a square-lattice spacing of $0.01$. 
Our results are qualitatively unchanged for $\theta=0.006$, with
identical exponents. This indicates that the value 
$\theta=0.012$ is well above $\theta_c$. The results presented below
were obtained in a system of size $512 \times 512$, and averaged over
$10$ (resp. $12$) runs in the case $\theta=0$ (resp. $\theta=0.012$).

To characterize the system in the absence of dynamical scaling of
correlations, we measure
the asymptotic (late-time) decay exponents $\beta_n$ of the
moments of the topological charge density distribution $P(|\rho|,t)$:
\begin{equation}
\label{EQ:beta}
\langle|\rho ({\bf r},t)|^n\rangle \sim t^{-\beta_n}\,,
\end{equation}
where $\langle\cdots\rangle$ denotes a spatial average. [We assume and
observe power-law decay with time.] The moment analysis
in terms of $\beta_n$ is analogous to multifractal analysis of 
static fractal \cite{Halsey86} or turbulent \cite{Frisch85} systems, 
only with the measuring length-scale replaced by the
inverse time elapsed since the quench (see, e.g., \cite{Roland89}). 

\begin{figure}
\epsfig{figure=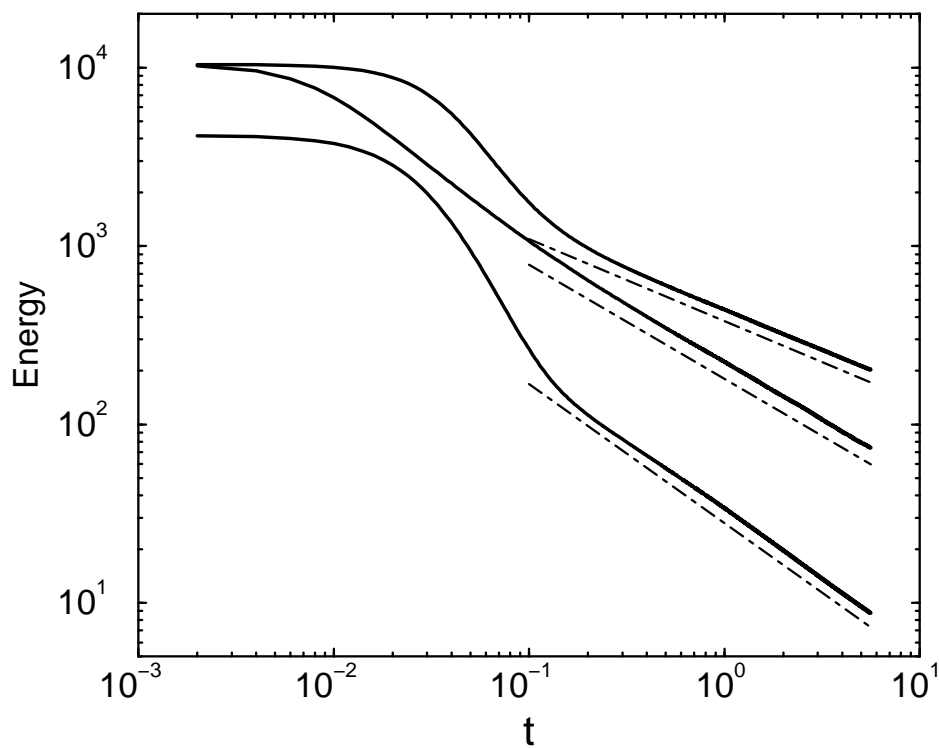,width=140mm,angle=270,silent=}
\caption{The exchange energy 
[first term of Eqn.~(\protect\ref{EQ:en})] for Skyrme-term amplitudes
 $\theta=0.012$ (upper curve) and $\theta=0$ (middle curve).
The lower curve shows the Skyrme energy [second term 
of Eqn.~(\protect\ref{EQ:en})] in the case $\theta=0.012$.
Even though the exchange energy dominates the total energy 
at late times, the decay exponent of the exchange energy depends on
the Skyrme-term amplitude $\theta$. The straight dot-dashed lines have slopes 
$-0.46$ (upper line), $-0.64$ (middle),
and $-0.78$ (lower).
\label{FIG:en}}
\end{figure}

The results for the asymptotic evolution of the energy density
$\epsilon$ are dramatically different for $\theta < \theta_c$ and 
$\theta > \theta_c$.  At late times, 
$\epsilon$ is dominated by the total number density of skyrmions 
and anti-skyrmions, so that $\epsilon \sim \langle |\rho| \rangle 
\sim 1/L_N^2 \sim t^{-2 \phi_N}$ and $\beta_1 = 2 \phi_N$.
Whereas systems with skyrmions unstable towards  shrinking 
($\theta< \theta_c$) have
$\phi_N = 0.32 \pm 0.01$, systems with $\theta > \theta_c$ have 
$\phi_N = 0.23 \pm 0.01$ , as shown in Fig.~\ref{FIG:en}. 
[Here we extract $\phi_N$ from the exchange energy,  
which dominates the total energy at late times (see Fig.~\ref{FIG:en}). 
The values extracted directly from $\langle |\rho| \rangle$ 
are $\phi_N = 0.31 \pm 0.01$ and $\phi_N = 0.21 \pm 0.01$, respectively, 
and are consistent. Error bars shown are always statistical errors.] 
To our knowledge this is the first example of a phase-ordering system 
in which a sub-dominant interaction controls the growth laws. 

\begin{figure}
\epsfig{figure=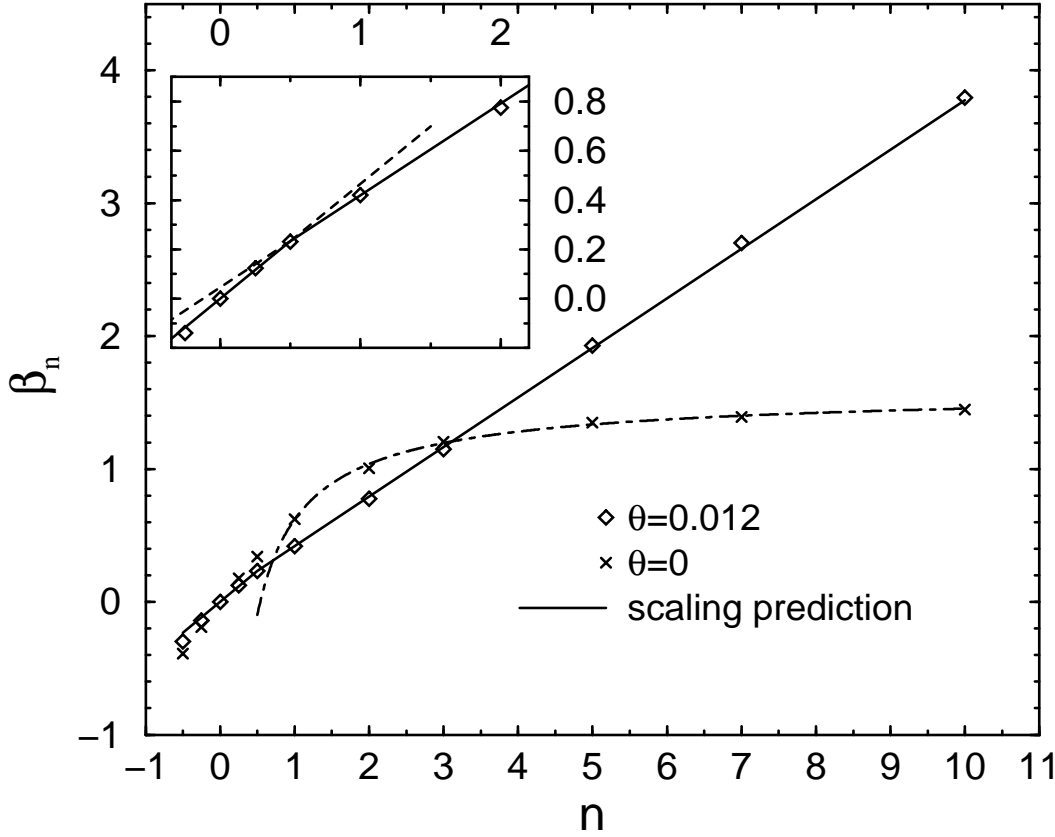,width=140mm,angle=270,silent=}
\caption{Growth exponents $\beta_n$ for the $n^{\rm th}$ moment of the
topological charge density, Eqn.~(\protect\ref{EQ:beta}). Shown are
the $\beta_n$ spectra for a system with Skyrme-term amplitude 
$\theta =0 <\theta_c$ and with $\theta=0.012 > \theta_c$. 
Error bars do not exceed the size of the symbols.
The full line shows the two--length-scale prediction of 
Eqn.~(\protect\ref{EQ:betanpred}) with $\phi_{\rm T}=0.187$ and 
$\phi_{\rm N}=0.210$.  The dotted-dashed line shows the phenomenological
fit to the $\beta_n$ spectrum in the $\theta=0$ case, given by 
Eqn.~(\protect\ref{EQ:empiricalbeta}) with $\phi_{\rm N} = 0.31$ and $p=0.78$.
Note that this fit is incompatible with the data for $n < 1$.
The inset highlights more clearly the discontinuity at $n=1/2$
in the slope of the $\beta_n$ curve for $\theta=0.012$, where the 
dashed lines are linear extrapolations of the regions above and below
$n=1/2$.
\label{FIG:betan}}
\end{figure}

\begin{figure}
\epsfig{figure=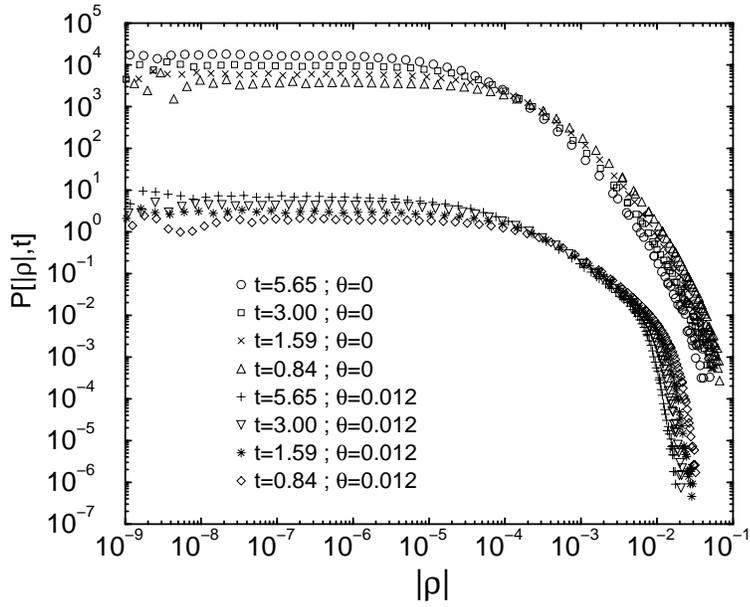,width=100mm,angle=270,silent=}
\caption{The distribution $P(|\rho|,t)$ of topological charge density
at indicated times $t$. The evolution of the small $\rho$ regime 
is well described by $P(0,t) \sim t^{0.75 \pm 0.05}$ in both cases. 
The vertical axis of the $\theta=0.012$ curves is offset by three 
decades for clarity.
\label{FIG:distrib}}
\end{figure}

Higher moments of the topological charge density, Eqn.~(\ref{EQ:beta}),
are shown in Fig.~\ref{FIG:betan}. The $\beta_n$ are multiscaling ---
they are not simply proportional to $n$.
This indicates that $P(|\rho|,t)$, shown in Fig.~\ref{FIG:distrib},
is determined by more than one length scale.  

We can explain many of the observed features of $P(|\rho|,t)$
using  the profile of a single Belavin-Polyakov (BP) skyrmion in 
combination with the multiple
length-scales $L_T$, $L_N$, and $L_C$. We first discuss the case 
$\theta>\theta_c$, where (after initial transients due to large gradients) 
skyrmions grow and annihilate with anti-skyrmions, 
and there is no unwinding through the lattice.

For $\theta>\theta_c$, a scaling argument involving both 
$L_T$ and $L_N$ suffices to characterize all of the $\beta_n$.
Picture the system divided into regions of size $L_N$, each containing a 
skyrmion or anti-skyrmion of size $L_T$ (see figures in 
\cite{Rutenberg95a,Zapotocky95}). Represent
each skyrmion as an isolated BP configuration of scale $L_T$,  so that 
$\rho({\bf r})=\pi^{-1} L_T^2/(L_T^2+r^2)^2$ \cite{Rajaraman82}
and $\langle |\rho({\bf r},t)|^n\rangle  \propto
{L_N}^{-2} L_T^{2 n} \int_0^{L_N} r\,dr\, (r^2 + L_T^2)^{- 2 n}$. For
$n>1/2$, the integral is dominated by the central region of radius 
$r \leq L_T$, and we obtain  
$\langle |\rho({\bf r},t)|^n\rangle \sim L_T^{2-2n} L_N^{-2}$. For $n<1/2$, on
the other hand, the integral is dominated by the large-$r$ regions,  giving 
$\langle |\rho({\bf r},t)|^n\rangle \sim L_T^{2n} L_N^{-4n}$. For
$n=1/2$, $\langle |\rho({\bf r},t)|^{1/2}\rangle \sim L_T L_N^{-2} \log(t)$.
In addition, the mixture of skyrmions and anti-skyrmions in the quenched 
system requires the presence of contours on which $\rho=0$ (probed by the 
third length, $L_c$, which is discussed later). 
We observe $P(|\rho|) = const$ for 
small $|\rho|$, which corresponds to a linear profile $\rho(z) \sim z$
with distance $z$ away from the $\rho=0$ contours. This constant regime
for small $\rho$ implies that moments with $n \leq -1$ diverge, and
the corresponding $\beta_n$ are not defined.  Summarizing: 
\begin{equation}
\label{EQ:betanpred}
\beta_n = 
 \left\{ \begin{array}{c}
 	2 n \phi_T + 2 (\phi_N - \phi_T), \ \ \ \ \ \ \ \ n \geq 1/2, \\
 	n (4 \phi_N - 2 \phi_T), \ \ \ \ \ \ \ \ \ \ \ \ \  \ -1 < n \leq 1/2.
 \end{array} \right.
\end{equation}
This piecewise-linear prediction agrees well with our data 
for $\theta > \theta_c$ in Fig.~\ref{FIG:betan}.  The best fit is 
obtained for the values $\phi_N=0.210 \pm 0.005$ and $\phi_T=0.18 \pm 0.02$
\footnote{These exponents are distinct, as can be seen from the distinct
linear regimes shown in the inset of Fig.~\protect\ref{FIG:betan}. In
light of this, there is covariance in the measured errors.}.

\begin{figure}
\epsfig{figure=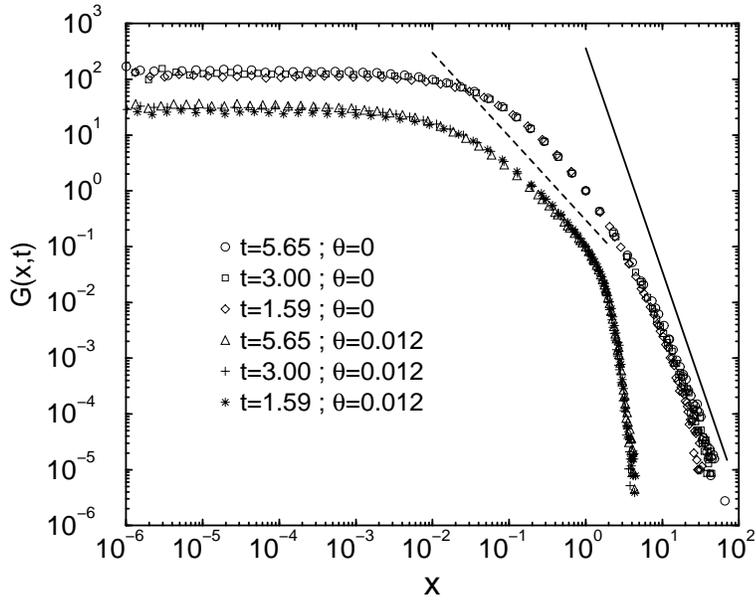,width=100mm,angle=270,silent=}
\caption{Re-scaled distributions $G(x,t)$ of topological charge density
at indicated times $t$. The vertical axis of 
the $\theta=0.012$ curves is offset by one decade for clarity.
The horizontal and vertical axes of the original $P(|\rho|,t)$ curves 
have been rescaled so that all curves collapse at $x=1$.
The straight lines drawn in the figure 
have slopes $-3/2$ (dashed line) and $-4$ (solid line).
\label{FIG:distribscaled}}
\end{figure}

For $\theta>\theta_c$, we show our data for the distribution 
$P(|\rho|,t)$ in Fig.~\ref{FIG:distrib} (lower set of curves).
The $P(|\rho|)=const$ regime is visible for small $|\rho|$, 
followed by a developing power-law regime for intermediate $|\rho|$ 
and an exponential cutoff at large $|\rho|$.
In Fig.~\ref{FIG:distribscaled}, the data from Fig.~\ref{FIG:distrib}
are rescaled to test the scaling ansatz $P(|\rho|,t) = f_1(t) G[|\rho| f_2(t)]$,
where $f_1$ and $f_2$ are functions of time only. A good quality of 
collapse is achieved for large and intermediate $x$, where $x= |\rho| f_2$. 
However, significant scaling violations are seen at small $x$. 

The distinct growth laws for $L_N$ and $L_T$ are sufficient to introduce scaling
violations in $P[|\rho|,t]$
\footnote{Additional scaling violations arise at small $x$ arise due to a
third length-scale $L_C$.}.
The approximate power-law behavior $G[x] \sim x^{-3/2}$ observed 
in the intermediate-$x$ range (see Fig.~\ref{FIG:distribscaled})
is consistent with a BP-like profile of the skyrmions. 
For a skyrmion of size $L_T$, 
$\rho({\bf r})=\pi^{-1} L_T^2/(L_T^2+r^2)^2$ implies 
the topological charge distribution of a single skyrmion
$P_{L_T}(\rho)={\sqrt{\pi} \over 2} L_T \rho^{-3/2}$ 
for $\rho \leq \rho_{max}= \pi^{-1} L_T^{-2}$. 
However, the tail of the BP profile is expected to be cut off at 
$r \geq L_N$ due to the presence 
of neighboring skyrmions and antiskyrmions; this corresponds to $\rho$ values
$\rho \geq \rho_c \sim t^{2 \phi_T - 4 \phi_N}$. 
Since $\rho_{max} / \rho_c \sim t^{4 (\phi_N - \phi_T)}$,  
the width of the $G[x] \sim x^{-3/2}$ regime is predicted to increase with $t$
--- hence violating the scaling ansatz.
Our data is consistent with this, and the lack of 
scaling of $P(|\rho|,t)$ is directly confirmed by the observed multiscaling 
character of $\beta_n$.

For $\theta<\theta_c$, the simple picture presented 
above does not apply. The instability of skyrmions towards unwinding in this
case leads to the presence of small skyrmions that significantly change 
$P(|\rho|,t)$
and hence $\beta_n$. Furthermore, we show below that unwinding 
skyrmions {\em cannot} be treated independently of annihilation.
To proceed, we consider the distribution density $H(R,t)$ of skyrmion 
sizes $R$ at time $t$
\footnote{To obtain Eqn.~(\protect\ref{EQ:betanpred}) we effectively used 
$H(R,t) =  \delta\left[ R-L_T(t) \right] /L_N^2$.}.

We see in Fig.~\ref{FIG:betan} that
$\beta_n$ saturate asymptotically at large $n$, for $\theta=0$.
This indicates the presence of small unwinding 
skyrmions, which dominate the moments of $P(|\rho|,t)$ 
for large enough $n$. Let us first assume that small skyrmions
(with sizes $R \ll L_T$)
do not participate in annihilation processes and unwind independently.
In that case, the size distribution function $H(R,t)$ satisfies
a continuity equation $\partial_t{H} + \partial_R J = 0$, 
where the ``size current'' is (see footnote $^{(1)}$) 
$J = H \partial_t{R} \sim -H/R^3$. To
maintain $\dot{N}$, where $N$ is the total number density of skyrmions and 
antiskyrmions, we must have a constant flux $J(R) = J(\xi)
\sim \dot{N}$ (where $\xi$ is the lattice scale),
or $H \sim - \dot{N} R^3$ for small $R$.  These small
skyrmions would dominate the high $\beta_n$, so that 
$\langle|\rho|^n \rangle \sim \int_\xi^{L_N} dR\,H(R) R^{2-2n} \sim \dot{N}
\xi^{6-2n} \sim t^{-(1+2 \phi_N)}$ for $n \geq 3$. 

For $\theta =0$, the measured $\beta_n$ do approach $1 +2 \phi_N$. This indicates
that unwinding is a significant relaxation process at late times.
However, they are not constant for large $n$ but instead fit the empirical form
\begin{equation}
\label{EQ:empiricalbeta}
	\beta_n = 1 + 2\phi_N - n^{-p}
\end{equation}
within error bars for all $n \geq 1$
(the unit prefactor of the power-law term ensures that $\beta_1 = 2\phi_N$).
In Fig.~\ref{FIG:betan}, we plot the best fit,
with $\phi_N = 0.31$ and $p=0.78$ 
($\phi_N=1/3$ and $p=2/3$ provide a comparable fit).
Our data for $P(|\rho|,t)$ are shown in Fig.~\ref{FIG:distrib} for
$\theta=0$ (upper set of curves). 
Similarly to the $\theta > \theta_c$ case, we see a $P(|\rho|) = const$
regime at low $|\rho|$.  As seen in Fig.~\ref{FIG:distribscaled}, 
$P(|\rho|,t)$ does not scale well, even for large $x$. 
An asymptotic power-law tail $P(|\rho|,t) \sim |\rho|^{-4}$ would
lead to $\beta_n$ saturating for $n \geq 3$.
The asymptotic approach {\em towards} that power-law (shown in 
Fig.~\ref{FIG:distrib}) is another manifestation 
of $\beta_n$ asymptotically {\em approaching} $1+2 \phi_N$ at large $n$.
From the previous paragraph, 
the existence of the non-zero correction term $n^{-p}$ for $n \geq 3$ 
is {\em inconsistent} with having a sub-population of
independently unwinding BP skyrmions. 

Two equivalent mechanisms can describe this breakdown
of the picture of small unwinding BP skyrmions for $\theta<\theta_c$. 
Either the profile of individual shrinking skyrmions is asymmetric and
hence not BP like, or the partial annihilation of skyrmions with 
anti-skyrmions changes the unwinding process \footnote{In the 
continuum limit, any non-BP structure in a skyrmion 
can be described by an additional anti-skyrmion component \cite{Belavin75}.}. 
In comparison, high-order gradient terms on their own do
not significantly affect the structure of an unwinding skyrmion until 
it approaches the lattice scale.

On the other hand, does unwinding affect the structure and evolution of 
annihilating skyrmions and anti-skyrmions?
To address this question, we need to probe the {\em large}-scale 
spatial distribution of the skyrmions and anti-skyrmions. 
We can do this by measuring the interface line-density of $\rho=0$. 
We measure this density, $L_C^{-1} \sim t^{-\phi_C}$, 
by counting the nearest neighbor lattice-sites between which 
$\rho$ changes sign, and find  $\phi_C = 0.41 \pm 0.01$ 
for both $\theta<\theta_c$ and $\theta > \theta_c$.
This exponent is significantly greater than $\phi_T$ 
or $\phi_N$, and is another demonstration of the strong
scaling violation seen in our system.

In addition to $L_C$, we have measured two other quantities that probe
the large-scale structure of the system. The low-$\rho$ value of the 
topological charge density $P(0,t) \sim 
t^{0.75 \pm 0.05}$ probes the charge distribution near 
the $\rho=0$ contours, while the half-width $L_{\rho} \sim 
t^{0.22 \pm 0.03}$ of $\rho-\rho$
correlations provides a measure of the typical skyrmion size.
The agreement of the given values
of the exponents of $L_C$, $P(0,t)$, and $L_{\rho}$ in the $\theta > \theta_c$ 
  and $\theta < \theta_c$ cases is striking
--- especially considering the dramatic differences in the $\beta_n$ spectrum.

We conclude simply: annihilation affects unwinding, but not vice versa.

In summary, we have studied phase-ordering dynamics in 
2D $O(3)$ systems supporting skyrmions.  By tuning the strength $\theta$ 
of the Skyrme term, we made isolated skyrmions
unstable to shrink and unwind ($\theta < \theta_c$) or to expand ($\theta 
> \theta_c$). We found dramatic changes in the multiscaling spectrum
$\beta_n$ of moments of the topological charge density between the two
cases. Having expanding skyrmions leads to a piecewise linear $\beta_n$ 
spectrum explained by taking into account that 
the average skyrmion size $L_T(t)$
and the average skyrmion separation $L_N(t)$ scale differently in time.
Having shrinking skyrmions leads to a curved $\beta_n$ 
spectrum remarkably well fit by Eqn.~(\ref{EQ:empiricalbeta}). 

ADR acknowledges support by EPSRC grant GR/J78044;  MZ by NSF grants
DMR-89-20538, DMR-94-24511, and DMR-95-07366.  We thank
Shivaji Sondhi for discussions.

\newpage

\end{document}